# From Play to Detection: Mini-SPACE as a Serious Game for Unsupervised Cognitive Impairment Screening

NANA TIAN*, Future Health Technologies, Singapore-ETH Center, Singapore

GIORGIO COLOMBO†, Future Health Technologies, Singapore-ETH Centre, Singapore and Chair of Cognitive Science, ETH Zurich, Switzerland

VICTOR R. SCHINAZI‡, Faculty of Society & Design, Bond University, Australia Future Health Technologies, Singapore-ETH Centre, Singapore

Early detection of Cognitive Impairment (CI) is critical for timely intervention, preservation of independence, and reducing the burden of dementia. Yet, most screening tools remain lengthy, clinic-based, and poorly suited for large-scale unsupervised deployment. This paper evaluates the test-retest reliability, validity, and usability of mini-SPACE, a short iPad-based serious game for detecting early signs of CI. Participants played mini-SPACE at home without supervision once a week for three weeks, with a longer version of the game in the final week. Mini-SPACE showed good test-retest reliability in unsupervised settings. While younger age was the primary predictor of performance, usability, and cognitive load, participants of all ages were able to complete the tasks and reported good usability and low cognitive load. Importantly, the prediction of scores in the Montreal Cognitive Assessment (MoCA) improved with repeated measures. These findings highlight mini-SPACE as a promising digital marker for scalable, age-sensitive screening and potential longitudinal tracking of CI.

## 1 Introduction

Cognitive impairment (CI) and dementia are rising worldwide, with prevalence projected to exceed 150 million cases by 2050 [74]. The associated direct and indirect costs of dementia are estimated to amount to hundreds of billions annually [96]. Beyond the financial impact, dementia erodes autonomy and undermines the ability of older adults to age successfully within their communities [81, 100]. Detecting CI early is critical for enabling timely interventions that preserve independence and improve quality of life [65]. Early identification allows clinicians to target modifiable risks, with several interventions shown to slow cognitive decline and reduce the incidence of dementia. For example, recent trials and large cohort studies have shown that interventions such as lowering blood pressure [43], use of GLP-1

Authors' Contact Information: Nana Tian, nana.tian@sec.ethz.ch, nana.tian@sec.ethz.ch, Future Health Technologies, Singapore-ETH Center, 1 College Ave E, CREATE Tower, Singapore; Giorgio Colombo, gicolombo@ethz.ch, Future Health Technologies, Singapore-ETH Centre, Singapore, Singapore and Chair of Cognitive Science, ETH Zurich, Zurich, Switzerland, gicolombo@ethz.ch; Victor R. Schinazi, vschinaz@bond.edu.au, Faculty of Society & Design, Bond University, Queensland, Australia and Future Health Technologies, Singapore-ETH Centre, Singapore, Singapore, vschinaz@bond.edu.au.

receptor agonists for diabetes [51], cessation of smoking [53], treatment of depression [48], and cataract surgery [60] are each associated with reduced risk. These findings highlight that identifying individuals at risk earlier creates meaningful opportunities to apply effective preventive measures and shift the trajectory of cognitive health at the population level.

Conventional paper-based assessments, such as the Montreal Cognitive Assessment (MoCA) [72] and the Mini-Mental State Examination (MMSE) [36], are widely used but require administration by trained personnel [16, 45], lack sensitivity to detect subtle changes in impairment, and their accuracy is often affected by language, culture, and education. Biomarkers such as Positron Emission Tomography (PET), Cerebrospinal Fluid (CSF), or blood tests can improve diagnostic accuracy, but require supervision and trained staff [82], and their cost and invasiveness mean they are most valuable once individuals at risk have already been identified through screening. Recent advances in digital technology have created new opportunities for cognitive screening outside of clinical and research settings [77, 79]. Digital tools offer several advantages over traditional methods, including automated administration and scoring, lower costs, scalability for large populations, and greater consistency across users by reducing human error and experimenter bias. Apps and serious games are particularly promising, as they provide an engaging framework that captures real-world cognitive demands such as memory, attention, and visuospatial abilities [31, 45].

Episodic memory has traditionally been the dominant focus in detecting CI and remains the primary target in many digital assessments, along with attention and executive function [45]. However, memory lapses and attention difficulties are also common in healthy ageing, making it difficult to separate expected changes from early disease [52]. One promising alternative is to probe spatial disorientation, which relies on navigation circuits in the entorhinal cortex and hippocampus known to be affected by tau and amyloid pathology in the early stages of CI [24, 91]. Digital navigation tasks may therefore provide a novel and biologically grounded avenue for detecting early signs of impairment that complements traditional memory and attention-based approaches.

Toward this end, Colombo et al. [19, 20] developed the Spatial Performance Assessment for Cognitive Evaluation (SPACE) as an iPad-based serious game that included a battery of spatial tasks to investigate differences in spatial ability indicative of CI. In recent years, SPACE has been progressively validated in multiple studies that provide converging evidence for its usability and utility as a cognitive assessment tool. Here, initial usability studies identified the most effective user interface and task configuration that ensures accessibility and reduces cognitive load in younger and older participants [20]. Performance in SPACE not only predicted MoCA scores beyond age, education, and risk factors, but also captured a distinct dimension of spatial navigation deficits associated with CI [22]. Furthermore, SPACE performance predicted hippocampal atrophy, a hallmark of Alzheimer's Disease (AD), in healthy participants after controlling for age and education [69]. Notably, SPACE was able to classify adjacent stages of clinically diagnosed CI with high sensitivity and specificity (AUCs 0.61 - 0.91), clearly distinguishing between healthy ageing, CI, and dementia [21].

Although SPACE has consistently shown robust performance and usability under supervised conditions [20–22], its validity and reliability in unsupervised settings, where differences in digital literacy, touchscreen familiarity, and task comprehension are critical, remain unknown. To address this limitation, we developed mini-SPACE, a time-efficient and accessible version designed for unsupervised use that includes the two most sensitive tasks (Visuospatial Training and Perspective Taking) from the six-task SPACE battery. Prior analyses by Colombo et al. [21] revealed that these tasks alone provided nearly the same sensitivity and specificity as the full assessment to discriminate between healthy, very mildly, and mildly impaired patients, with completion times around 11 minutes. Based on this evidence, this study evaluates mini-SPACE to determine whether its reliability, validity, and usability are preserved when deployed unsupervised outside research settings and the clinic over a three-week period. We further examine how performance and usability vary between age and gender, and compare results between unsupervised and previously recorded supervised assessments. Prior work in spatial navigation consistently reports robust gender-related differences[10, 71, 73], and both age and gender have also been shown to shape technology perception and interaction patterns in HCI[17, 49]. Therefore, we balanced age and gender in our usability analysis to align with evidence from both navigation and HCI research and to enable clearer interpretation of demographic effects. Together, these analyses aim to establish whether mini-SPACE can serve as a scalable digital tool for early cognitive screening under unsupervised real-world conditions.

## 1.1 Research Questions

The research questions of this study are therefore as follows:

- Q1: Does performance in mini-SPACE demonstrate acceptable test-retest reliability when used in an unsupervised setting?
- Q2: Can mini-SPACE predict CI as assessed by scores in the MoCA?
- Q3: Does mini-SPACE under unsupervised conditions meet or exceed established usability benchmarks from the System Usability Scale (SUS), User Experience Questionnaire (UEQ), and NASA Task Load Index (NASA-TLX)?
  - Q3.1: Does usability in mini-SPACE change across the three weeks of unsupervised testing?
- Q4: How does performance in mini-SPACE vary across age, gender, and repeated testing sessions?
- Q5: Are there performance differences across age groups and gender when comparing the supervised (SPACE) and unsupervised (mini-SPACE) assessments?
- Q6: Are there usability differences across age groups and gender when comparing the supervised and unsupervised assessments?

## 1.2 Contributions

The contributions of this work are fourfold:

- **A novel unsupervised digital assessment of CI.** We present mini-SPACE, a short iPad-based serious game designed for home-based cognitive screening that uses deficits in navigation ability as an indicator of CI.
- **Comprehensive evaluation of reliability, validity, and usability**. Through a three-week multi-session study with 93 participants, we demonstrate strong test-retest reliability, robust associations with MoCA, high usability, and low cognitive workload scores, providing empirical evidence for effective interaction design for older adults.
- **Comparison with supervised studies and demographic insights.** By contrasting mini-SPACE with prior supervised studies and examining effects of age and gender, we show that unsupervised administration achieves comparable validity and usability, while emphasising the importance of adaptive, user-centred support for diverse populations.
- **A reusable data processing platform.** We designed and developed the SPACE Data Parser, a novel website that automatically transforms raw game data from SPACE into standardised CSV outputs with calculated parameters, advancing open, reproducible research practices within the Human Computer Interaction and digital health communities.

## 2   Related Work

### 2.1   Designing for Ageing Populations

Digital approaches leverage the widespread availability of computers, tablets, and smartphones to deliver objective, scalable health assessments in everyday environments [31, 45]. Here, apps can support frequent testing, reduce the costs associated with travelling, and empower older adults to take a more active role in managing their health [3]. Although smartphone and tablet ownership among older adults has grown in recent years [34], many still face barriers related to reduced motor precision, sensory decline, and technology-related anxiety [28, 37]. These obstacles often manifest in frustration and disengagement, contributing to low retention rates [87]. Designing tools for this group requires a focus on clarity and simplicity [38]. Gamification and serious games offer a promising solution by embedding training and guidance directly into the assessment experience. Through simplified controls, interactive tutorials, and clear instructions, gamified tasks can gradually familiarise users with the app while reducing technological anxiety [4, 7]. In addition, features such as storytelling and progressive challenges can scaffold learning and sustain engagement during repetitive activities that are essential for health assessments [41].

Over the last decades, a variety of tools have been developed for the supervised and unsupervised assessment of CI [32, 44, 77, 89, 90, 95]. A recent review of 23 remote, unsupervised digital tools [79] reported high feasibility (consent 74-93%, adherence 63-94%), acceptable reliability (ICCs 0.65-0.95), and moderate validity for detecting preclinical AD (AUC 0.63-0.83). While these results are encouraging, most existing tools offer focus on deficits in episodic memory, processing speed, executive function, and language [79]. This is surprising, given that deficits in spatial orientation and navigation are a known marker of hippocampal dysfunction and one of the earliest behavioral manifestations of CI [2, 20, 23, 47, 67, 91]. Moreover, the few tools that target visuospatial and navigation abilities typically assess static tasks such as image-based recognition [8, 46] or object location [55], rather than the dynamic spatial and navigational skills needed for real-world movement and orientation.

### 2.2   Navigation Tools for Detecting Cognitive Impairment

Researchers in spatial cognition have developed a wide variety of tasks to measure navigation ability, many of which have been successfully adapted to virtual environments and used to assess deficits associated with healthy and pathological ageing [9, 47, 62, 76, 86, 98]. For example, the *Virtual Supermarket Test* (VST) can be delivered supervised or unsupervised and is used to assess participants' spatial memory and executive function by requiring them to recall items, plan routes, and orient within a virtual supermarket [11]. Although effective in detecting CI, validation of the VST is largely confined to research settings [91], and broader applicability may be limited by side effects such as simulator sickness. Adopting a serious game format, *Sea Hero Quest* (SHQ), the *Leiden Navigation Test* (LNT), and the *Virtual Environments Navigation Assessment* (VIENNA) each target different aspects of navigation ability in a more engaging game format. SHQ [27] is an unsupervised assessment that combines wayfinding (e.g., map interpretation, route planning, spatial memory) and path integration (e.g., egocentric updating and working memory) tasks. SHQ has shown good ecological validity with in-game performance correlating with real-world navigation [26], but its test-retest reliability was found to be inconsistent [25, 83]. Moreover, SHQ has yet to be validated for detecting CI beyond associations with genetic predisposition [24]. In contrast, the LNT [15, 94] is a supervised assessment, in which participants watch a video of a route and are tested on their recall and memory for landmarks. To date, the LNT has only been deployed in healthy populations and assessed relative to self-report navigation questionnaires, rather than established cognitive screening tools. More recently, VIENNA [80] was introduced as a passive, easy-to-administer assessment to complement other neuropsychological tests. VIENNA has demonstrated a stable correlation with the MMSE, but its passive nature can lead to disengagement over time. Despite these advancements, none of the existing tools have been rigorously evaluated in terms of their reliability, validity, and usability in unsupervised contexts, which are critical for scalable, home-based screening [91].

## 3 Methodology

### 3.1 Participants

We recruited 108 participants balanced across gender and age groups, with support from a dedicated Telegram group specifically established to expose our study advertisements to nearly 20,000 people. Previous work using SPACE has demonstrated robust age-related effects with medium to large effect sizes on usability and performance outcomes [20]. Assuming $\alpha$ = .05 and $1 - \beta$ = .80, a priori power analyses indicated that medium to large effects ($f^2$ = 0.15 - 0.35; $f$ = 0.25 - 0.40) could be detected with 77 participants for a hierarchical linear regression with three predictors, 72 participants for a mixed factorial ANOVA, and 108 participants for a between-groups factorial ANOVA that we report in this paper, respectively. Following exclusion of participants with protocol violations (e.g., repeated gameplay within the same week) or incomplete data, the final dataset included 93 participants (45 male, 48 female) ranging in age from 20 to 79 years (median = 52.0; mean = 48.9, SD = 16.96). All participants had normal or corrected-to-normal vision and no psychiatric or neurological diagnoses. Table 1 presents the demographic characteristics of the participants. Each experimental session lasted between 5 and 15 minutes, and participants were compensated with $135 regardless of their performance. Informed consent was obtained from all participants prior to the commencement of the experiment. All experimental protocols were approved by the National Ethics Commission, and the experiment was performed in accordance with the Declaration of Helsinki.

Table 1. Demographic characteristics and summary of measures in the mini-SPACE study.

| Age Group | Gender | N | Mean | SD | Min | Max |
|---|---|---|---|---|---|---|
| 20–40 | Male | 15 | 29.73 | 6.92 | 22 | 40 |
| | Female | 16 | 27.19 | 5.38 | 21 | 37 |
| 41–60 | Male | 15 | 50.40 | 5.91 | 42 | 59 |
| | Female | 16 | 51.25 | 5.61 | 42 | 60 |
| 61–90 | Male | 15 | 66.86 | 5.39 | 61 | 79 |
| | Female | 16 | 67.00 | 5.13 | 61 | 79 |
| MoCA score | % Healthy/Impaired | 70% / 30% | | | | |
| Gender | % Male/Female | 47% / 53% | | | | |
| Education | % High school/University | 31% / 69% | | | | |
| Handedness | % Right/Left | 90% / 10% | | | | |
| Vision defects | % Yes/No | 46% / 54% | | | | |
| Chronic Health conditions | % Yes/No | 22% / 78% | | | | |
| Navigation training | % Yes/No | 11% / 89% | | | | |

### 3.2 Materials

Participants completed an onboarding session at the laboratory, during which they filled out a sociodemographic and health questionnaire and the MoCA [72]. They then took part in an unsupervised phase at home, consisting of three weekly sessions where they completed the mini-SPACE assessment and usability assessments comprising the SUS [14], the UEQ [59], and the NASA-TLX [40] after each session.

### 3.2.1 Questionnaires.

**Demographics and Health:** The sociodemographic and health questionnaire collected information on age, gender, education, handedness, tablet use, and navigation training. The instrument also included questions on health status and habits (i.e., vision, chronic conditions, brain injury, smoking, alcohol use, sleep, physical activity) as well as psychosocial well-being (i.e., self-reported depression, anxiety, stress).

**MoCA:** The MoCA is a 30-point in-person screening tool for screening CI that evaluates memory, executive function, visuospatial abilities, language, attention, and orientation. Scores below 26 indicate mild impairment. The Moca has a sensitivity of 90% and specificity of 87%.

**SUS:** The SUS is a 10-item questionnaire yielding scores from 0 to 100, with higher values reflecting better usability. The SUS has demonstrated strong reliability ($\alpha$ = 0.85-0.90) and sensitivity to product differences [5].

**UEQ:** The UEQ is a 26-item questionnaire measuring user experience across six dimensions on bipolar scales from −3 to +3, with higher values indicating more positive evaluations. Subscale scores for attractiveness, pragmatic, and hedonic qualities were computed following standardised procedures [59]. The reliability of the UEQ ranges from $\alpha$ = 0.55-0.95 [59].

**NASA-TLX:** The NASA-TLX is a 6-item questionnaire that measures perceived cognitive workload across multiple domains. Total scores range from 0 to 100, with higher values indicating greater workload. The NASA-TLX shows good reliability for both individual items and overall scores ($\alpha$ > 0.80) [99].

**Debriefing:** This debriefing questionnaire invites participants to provide feedback on their experience with the mini-SPACE game. It focuses on clarity of instructions, ease of controls in both the training and perspective taking tasks, and suggestions for improving enjoyment and long-term motivation.

### 3.2.2 Hardware.

Since this was an unsupervised assessment, participants had the choice to install SPACE on their own device or borrow one from the lab. Lab devices included 10.2-inch iPads and 11-inch iPad Pros, all equipped with Wi-Fi but varying storage capacities. A minimal iOS version of 10.0+ was required for all devices. To balance temporal precision and system performance, the behavioural data were sampled at a frequency of 4 Hz, which was deemed optimal for capturing meaningful input while maintaining game responsiveness. Questionnaire responses were collected in person using the Qualtrics XM online platform (www.qualtrics.com) on a MacBook Pro, and remotely at participants' homes via the same platform using the iPad.

### 3.2.3 Software: Mini-SPACE.

SPACE is a serious game developed for the iPad that provides an in-depth evaluation of spatial and navigation abilities to identify deficits that may signal early signs of CI. The game is set in a virtual environment resembling an alien planet and includes five core spatial tasks: Visuospatial Training, Path Integration, Egocentric Pointing, Mapping, and Perspective Taking. Details of SPACE and the tasks have been described in earlier work by Colombo et al. [20, 21] and Minta et al. [69]. Mini-SPACE is an adapted version of the game that includes only the Visuospatial Training and perspective tasks, allowing for faster unsupervised deployment, see Figure 1. Below, we expand on how these two tasks were tailored and implemented for the current study.

**Visuospatial Training:** This task consists of two phases (Rotation and Movement) designed to assess basic visuospatial skills. In the Rotation phase, participants swipe the screen to follow the various rotations performed by a robot guide. In the Movement phase, participants first follow the robot forward by tapping and holding a button on the screen until they reach the robot's position, and then repeat the rotation task they performed earlier. For Weeks 1 and 2, the rotation angles are drawn from a small set: 90°, 45°. These values are then randomly assigned a direction, clockwise or counter-clockwise, by multiplying each by either +1 or -1, creating sequences like 90°, -45°,45°, -90°, depending on the randomised signs. Since every rotation is followed by a return to the original orientation, angles typically appear in opposing pairs (e.g., 45°, -45°). For Week 3, the angle set becomes more comprehensive, covering the full 360° circle in 45° increments: 45°, 90°, 135°, 180°, 225°, 270°, 315°. As with the earlier weeks, each angle is randomly assigned a direction (positive or negative) to simulate clockwise or counterclockwise rotation.

**Perspective Taking:** The Perspective Taking task assesses the participants' ability to judge the spatial relationship between landmarks from various perspectives. Over a series of trials, participants consult a top-down map of a virtual environment and are asked to identify the direction of a landmark by imagining themselves standing at a specified location and facing a given orientation (e.g., "imagine you are standing at the rocket, facing the tree, point to the cave). For Weeks I and II, the map contained four landmarks and participants were tested over six trials with different combinations of imagined positions, orientations and target landmarks. In Week 3, the map is expanded to seven landmarks and sixteen trials.

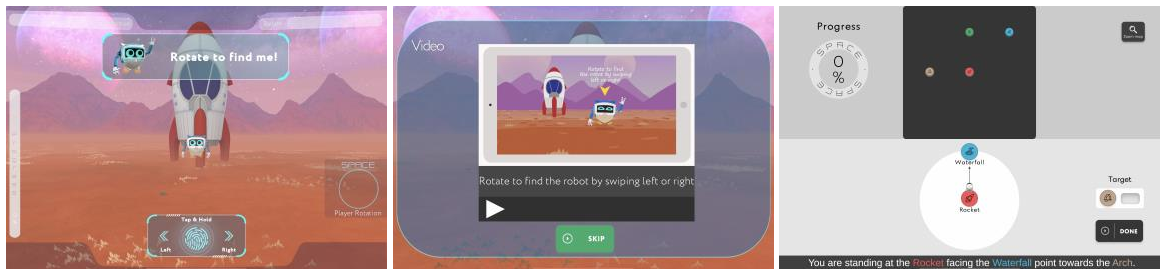

(a) Rotation Task: Participants swipe to ro- tate the viewpoint to follow the robot's di- rection until the robot is successfully cap- tured inside a crosshair.

(b) Video Tutorial for the Visuospatial Training task: Participants can access video tutorials for training on both Rotation and Movement tasks.

(c) During Perspective Taking Week 1 and 2, the map features four landmarks and is used across six trials.

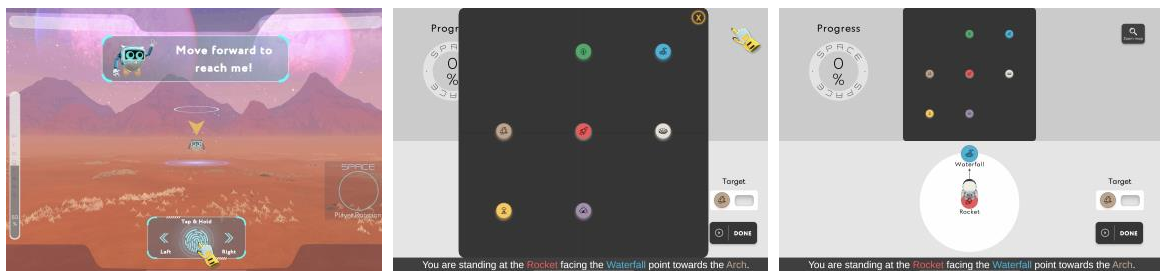

(d) Movement Task: To move forward, par- ticipants must tap and hold the button until they reach the circular target positioned in front of the robot.

(e) Animated Tutorial for the Perspective Taking task: When entering the scene, the animation automatically plays at a slow pace.

(f) During Perspective Taking Week 3, the map features seven landmarks and is used across sixteen trials.

Fig. 1. Screenshots from mini-SPACE include the training task with demonstration of Rotation and Movement (a, d), video and animated tutorials for the various tasks (b, e), and maps for the Perspective Taking for different weeks (c, f)

### 3.3 Procedure

The study consisted of an onboarding session with the experimenter at the laboratory, followed by three unsupervised home sessions once per week over three consecutive weeks. Upon arriving at the laboratory, a member of the study team conducted a face-to-face interview in English to explain the study procedures and obtain voluntary written informed consent. Following this, participants completed the demographics and health questionnaire and the MoCA, administered by a certified team member. Next, the experimenter assisted participants in downloading the mini-SPACE app on the participant's iPad and verified that the app was properly installed and functioning. For participants who did not own an iPad, a device was temporarily provided upon signing a loan agreement. The three home sessions consisted of completing the two tasks in mini-SPACE, followed by the SUS, UEQ, NASA-TLX, and debriefing questions delivered via Qualtrics. Figure 2 presents a schematic of the experimental flow.

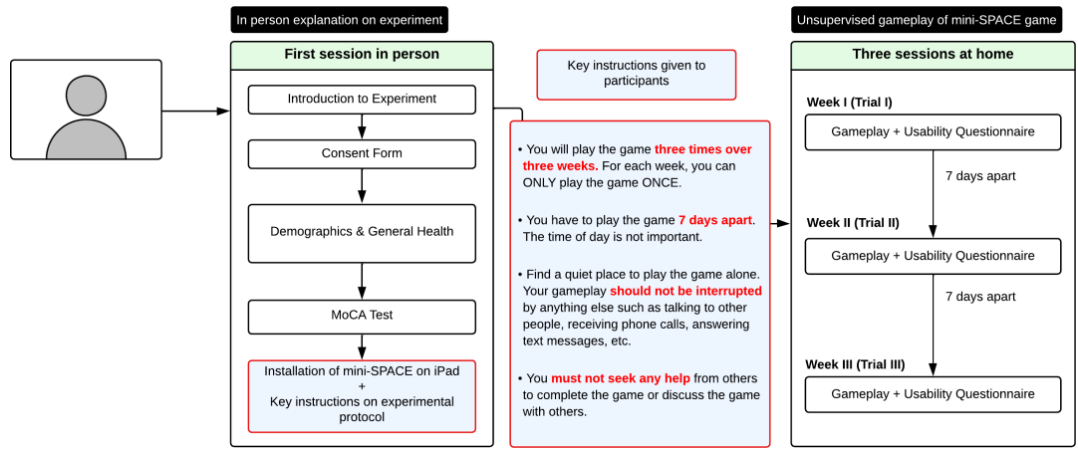

Fig. 2. A schematic of the experiment flow for the unsupervised evaluation of mini-SPACE, including key instructions given to the participants (on-site) and weekly gameplay sessions followed by usability questionnaires (home-based).

### 3.4 Design and Analysis

During gameplay, data such as player position, landmark coordinates, time stamps, and angular error are saved in JSON format. We developed a dedicated web platform, the SPACE Data Parser, to standardise data extraction and metric computation across datasets, ensuring consistency and reproducibility. Researchers can access the website via the link[1].

*3.4.1 SPACE Data Parser.* Figure 3 illustrates the interface of the SPACE Data Parser, where users can access a detailed user manual via the tag-style button located at the top left corner of the webpage. Upon first opening the webpage, they are prompted to upload their data via a drag-and-drop interface or by selecting a file directly from their local computer or iPad. Both a single JSON file and compressed .zip archives are supported, with the platform tested to successfully process zip archives containing over 400 JSON files. Data export is automatically adapted to the researcher's configuration, with two available options: (i) Detailed Measures, which outputs task-level data only for the tasks selected in the configuration (e.g., if the perspective taking task is not included, its variables will not appear), and (ii) Quick Summary of Measures, which provides a concise overview of the main outcomes. After selecting an option and uploading the file(s), the right-hand panel displays a dynamically generated list of available variables grouped by category (e.g., Player information, Training, Perspective taking). Researchers can flexibly select or deselect specific columns for download according to their analytic needs. Once the selection is made, clicking the Download button initiates the export process, which typically completes within one second even for large datasets.

---
[1] https://space.fhtsecethz.org

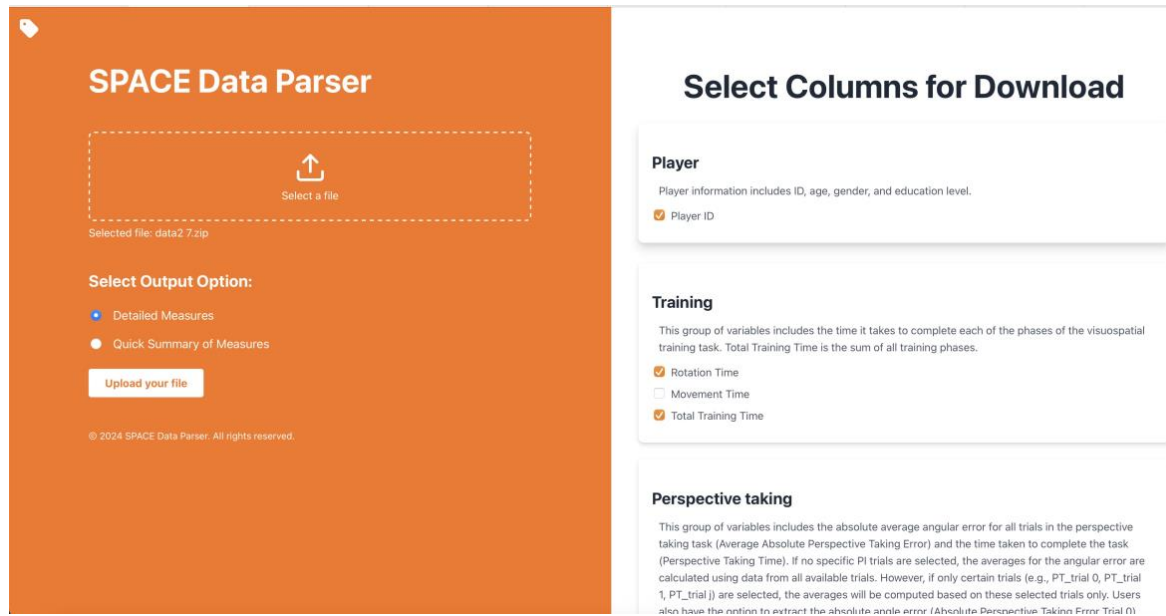

Fig. 3. A screenshot of SPACE data parser website

*3.4.2 Quantitative analysis.* For both the Visuospatial Training and Perspective Taking tasks, we extracted several performance parameters following the methodology outlined by Colombo et al. [20]. Here, we computed total training time as the combined duration (in seconds) required to complete both the Rotation and Movement tasks. We also considered the duration of each task separately to further differentiate their sensitivity. For the Perspective Taking task, we calculated Perspective Taking error, defined as the average angular deviation (in degrees) across trials between the participant's estimated direction and the actual target landmark. We also created a composite SPACE error score by standardising (z-scoring) and then averaging the results from the Visuospatial Training and Perspective Taking tasks. The scores for the post-game usability questionnaires were computed according to their respective standardised procedures [14, 40, 59]. For the majority of the analyses, we used scores from the three weeks. However, we focused only on Week 1 for the analysis that compared results from mini-SPACE to previously collected supervised SPACE data.

Prior to conducting inferential statistics, we verified whether our data violated the assumptions of normality of residuals and homoscedasticity in correlations, t-tests, Analysis of Variance (ANOVA), and linear regression. When assumptions were met, parametric statistics were used. When assumptions were violated, we used non-parametric statistics. For all ANOVA models, age was treated as a categorical factor with three groups: younger adults (20-40), middle-aged (41-60), and older adults (61-90). All post hoc tests were adjusted for multiple comparisons using Holm-Bonferroni correction. Regarding the effect size, for variables with two levels (e.g., Gender, Supervision), we report effect sizes using Cliff's delta ($\delta$). For variables with three or more independent groups (e.g., Age.group), we report epsilon squared ($\epsilon^2$) from Kruskal–Wallis tests. For within-subjects factors with three or more levels (e.g., Week), we report Kendall's $W$ for the overall effect and rank-biserial correlations ($r_{rb}$) for post hoc contrasts. For one-sample Wilcoxon signed-rank tests against benchmarks, we report effect sizes as correlation coefficients ($r$).

To address Q1, we employed Spearman's rank-order correlation to evaluate the test-retest reliability of the standardised SPACE error across the three unsupervised testing sessions. Then, we used a one-way Aligned Rank Transform (ART) ANOVA to examine changes in performance with the standardised SPACE error. To address Q2, we conducted separate hierarchical linear regression models to examine whether the standardised SPACE error or task-specific measures predicted MoCA scores, controlling for age, gender, and education. For Q3, we used separate Wilcoxon signed-rank tests to compare usability ratings (SUS, UEQ subscales, NASA-TLX) against their established benchmarks across the three weeks. For Q3.1, we used a one-way ART ANOVA to examine changes in usability rating (SUS, UEQ, NASA-TLX) across weeks. To address Q4, we used a 2 (Gender) × 3 (Age Group) × 3 (Week) mixed ART ANOVA to examine demographic differences in the standardised SPACE error, with Week included as a within-subject factor. To address Q5, we used a pooled dataset from previous supervised SPACE studies (uSPACE, targetSPACE, bioSPACE, trainSPACE; see Table 4) conducted by Colombo et al. [20, 22] and Minta et al. [69] and compared it to Week 1 of mini-SPACE. For comparability, we recalculated supervised performance metrics using only the Visuospatial Training and Perspective Taking tasks, and analyses were restricted to the supervised configurations that matched this condition (targetSPACE, bioSPACE, trainSPACE, and the simplified condition in uSPACE). Here, a 2 (Gender) × 2 (Supervision) × 3 (Age Group) factorial ART ANOVA was used with the standardised SPACE error as the dependent variable. Finally, for Q6, we used separate 2 (Gender) × 2 (Supervision) × 3 (Age Group) factorial ART ANOVAs with scores from the SUS, UEQ, and NASA-TLX as dependent variables. For comparability, only the simplified condition from the uSPACE was included in the supervised dataset for this analysis.

*3.4.3 Qualitative analysis.* To complement quantitative measures, we asked participants to provide open-ended feedback on their experience with the mini-SPACE tasks. Specifically, responses were reviewed and grouped into themes relating to clarity of instruction, usability, and engagement. For each theme, we report the proportion of participants who endorsed it, together with descriptive breakdowns by gender and age group, and include representative quotes to illustrate these perspectives.

## 4 Results

### 4.1 Q1: Test-retest Reliability

Descriptive statistics for the mini-SPACE tasks and self-report usability questionnaires are presented in Table 2. Results revealed good test-retest reliability, with strong positive correlations between adjacent weeks (Weeks 1 and 2, $\rho = 0.70$, $p < .001$; Weeks 2 and 3, $\rho = 0.75$, $p < .001$), and a weaker but still robust correlation between Weeks 1 and 3 ($\rho = 0.60$, $p < .001$). For completeness, we also calculated Intraclass Correlation Coefficients (ICC), see Figure 4. Here, results indicated good reliability for single administrations of SPACE (ICC$_{(2,1)}$ = 0.67, 95% CI [0.58, 0.74]) and excellent reliability when scores were averaged across the three sessions (ICC$_{(2,k)}$ = 0.86, 95% CI [0.81, 0.89]; where $k$ = 3 sessions). Variance decomposition showed that 67% of the variability was attributable to stable between-person differences, while 33% reflected within-person error (residual error).

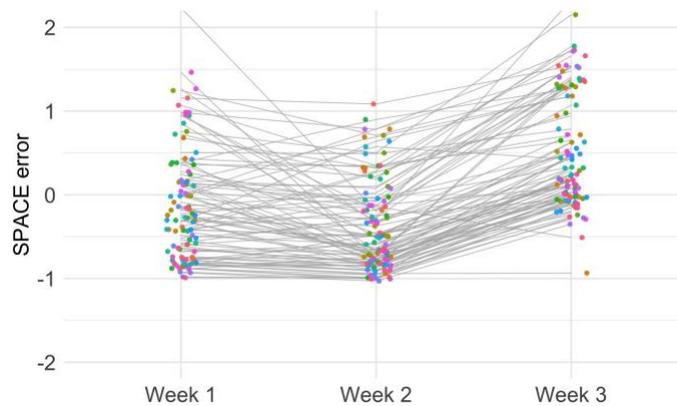

Fig. 4. Spaghetti plot showing performance across weeks, with participants generally maintaining their relative standing).

*4.1.1 Performance changes across weeks.* Results of the ART ANOVA on SPACE error revealed a significant effect of Week, $F(2, 184) = 140.07$, $p < .001$; $W = 0.61$. Post hoc comparisons showed that SPACE errors were higher in Week 3 than Week 1 ($p < .001$, $r_{rb} = 0.72$) and Week 2 ($p < .001$, $r_{rb} = 0.95$), and errors in Week 1 were also significantly higher than in Week 2 ($p < .001$, $r_{rb} = 0.48$).

## 4.2 Q2: Validity with MoCA

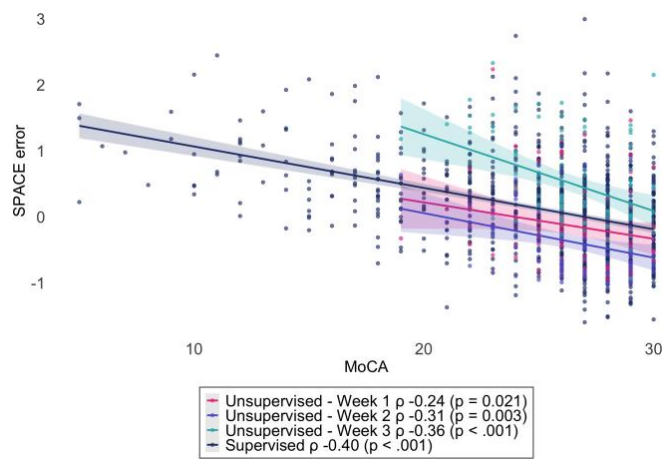

Fig. 5. This figure illustrates the associations between MoCA scores and SPACE errors across three weeks in the unsupervised condition, alongside supervised results from previous studies.

To examine whether SPACE error predicted MoCA scores, we tested three linear regression models extending from a common demographic baseline (age, gender, education) model for each of the testing weeks (Table 3). Results revealed that the baseline model was not significant, $F(3, 89) = 2.26$, $p = .087$, $R^2 = .07$, indicating that demographics alone could not explain variance in MoCA scores. Model 1 (baseline + Week 1 SPACE error) was also not significant, $F(4, 87) = 1.96$, $p = .107$, $R^2 = .08$. Week 1 SPACE error showed a negative but not significant association with MoCA ($\beta = -0.51$, $p = .177$). Model 2 (baseline + Week 2 SPACE error) was significant, $F(4, 88) = 4.04$, $p = .005$, with $R^2 = .16$. Here, higher SPACE error scores in Week 2 significantly predicted lower MoCA scores ($\beta = -1.07$, $p = .004$). Model 3 (baseline + Week 3 SPACE error) was also significant, $F(4, 88) = 5.45$, $p < .001$, with $R^2 = .20$. In Week 3, being female predicted higher MoCA scores ($\beta = -0.99$, $p = .034$), and SPACE error remained a strong negative predictor of MoCA scores ($\beta = -1.24$, $p < .001$). Figure 5 presents a scatterplot of SPACE error and MoCA scores for each of the weeks.

Table 2. Descriptive statistics of gameplay and usability (Median, Mean ± SD) by Age Group across Week 1-3

| Condition | | Week 1 | Week 2 | Week 3 |
|---|---|---|---|
| **Gameplay Outcomes** | **Age** | **Median, Mean ± SD** | | |
| Rotation Time | Young | 28.12, 28.32 ± 4.75 | 24.06, 24.01 ± 2.19 | 91.51, 94.70 ± 33.22 |
| | Middle-aged | 29.61, 32.58 ± 9.77 | 27.26, 30.51 ± 9.53 | 101.90, 106.97 ± 30.31 |
| | Old | 37.20, 59.39 ± 62.97 | 33.19, 36.15 ± 11.42 | 101.71, 111.31 ± 27.15 |
| Movement Time | Young | 37.61, 39.25 ± 5.74 | 35.89, 36.52 ± 3.93 | 78.12, 81.67 ± 16.52 |
| | Middle-aged | 46.47, 50.76 ± 13.90 | 42.21, 44.82 ± 8.91 | 96.60, 101.68 ± 22.14 |
| | Old | 50.02, 52.96 ± 9.85 | 48.95, 50.40 ± 8.99 | 108.28, 114.17 ± 29.41 |
| Total Training Time | Young | 67.06, 67.57 ± 8.37 | 59.43, 60.53 ± 5.00 | 171.10, 176.37 ± 48.05 |
| | Middle-aged | 76.03, 83.33 ± 20.78 | 70.31, 75.34 ± 14.46 | 202.22, 208.65 ± 43.58 |
| | Old | 99.49, 112.36 ± 63.87 | 83.40, 86.55 ± 17.48 | 212.54, 225.48 ± 38.56 |
| Perspective Taking Error | Young | 17.61, 28.11 ± 25.27 | 15.20, 22.66 ± 20.58 | 14.14, 19.39 ± 17.58 |
| | Middle-aged | 32.86, 36.62 ± 25.26 | 21.38, 34.03 ± 28.50 | 22.47, 32.11 ± 27.44 |
| | Old | 51.83, 60.27 ± 29.78 | 41.69, 43.93 ± 27.12 | 38.29, 39.08 ± 25.20 |
| SPACE error | Young | -0.36, -0.35 ± 0.53 | -0.41, -0.38 ± 0.49 | -0.19, -0.15 ± 0.59 |
| | Middle-aged | 0.04, 0.06 ± 0.61 | 0.08, 0.06 ± 0.57 | 0.11, 0.10 ± 0.64 |
| | Old | 0.74, 0.74 ± 0.62 | 0.52, 0.51 ± 0.54 | 0.42, 0.43 ± 0.66 |
| **Usability Outcomes** | **Age** | **Median, Mean ± SD** | | |
| SUS Score | Young | 75.00, 75.58 ± 13.34 | 80.00, 79.27 ± 11.44 | 77.50, 77.66 ± 11.83 |
| | Middle-aged | 72.50, 69.91 ± 15.97 | 77.50, 75.52 ± 16.53 | 77.50, 75.43 ± 14.32 |
| | Old | 62.50, 62.12 ± 17.20 | 67.50, 66.59 ± 14.13 | 70.00, 68.52 ± 15.64 |
| NASA-TLX | Young | 19.17, 20.67 ± 15.09 | 11.67, 15.59 ± 13.10 | 15.83, 20.54 ± 16.20 |
| | Middle-aged | 19.17, 23.28 ± 18.30 | 17.50, 20.22 ± 17.80 | 15.83, 19.94 ± 17.66 |
| | Old | 36.67, 34.45 ± 14.19 | 27.50, 28.59 ± 16.03 | 28.33, 29.70 ± 15.71 |
| UEQ Attractiveness | Young | 0.00, -0.09 ± 0.66 | 0.00, -0.05 ± 0.52 | 0.00, -0.08 ± 0.37 |
| | Middle-aged | -0.17, -0.16 ± 0.49 | 0.00, -0.02 ± 0.47 | 0.00, -0.13 ± 0.44 |
| | Old | 0.00, 0.13 ± 0.55 | 0.00, 0.13 ± 0.50 | 0.00, 0.01 ± 0.46 |
| UEQ Hedonic | Young | 0.00, -0.05 ± 0.62 | 0.00, -0.07 ± 0.49 | 0.00, -0.09 ± 0.42 |
| | Middle-aged | -0.17, -0.11 ± 0.46 | 0.00, -0.07 ± 0.45 | 0.00, -0.15 ± 0.42 |
| | Old | 0.00, 0.12 ± 0.54 | 0.00, 0.10 ± 0.47 | 0.00, 0.03 ± 0.47 |
| UEQ Pragmatic | Young | 0.00, -0.11 ± 0.63 | 0.00, -0.04 ± 0.53 | 0.00, -0.07 ± 0.38 |
| | Middle-aged | -0.17, -0.18 ± 0.50 | 0.00, -0.05 ± 0.47 | 0.00, -0.12 ± 0.45 |
| | Old | 0.00, 0.13 ± 0.55 | 0.00, 0.14 ± 0.51 | 0.00, 0.01 ± 0.45 |

\* Participants were classified into three age groups: 20-41 years as **Young** group, 41-60 years as **Middle-aged** group, and 61-90 years as **Old** group.

To further examine which mini-SPACE tasks best predict MoCA performance, we conducted a series of regression models using demographic covariates along with task-specific measures from Week 1, Week 2, and Week 3. The Week 1 model was not significant, $F_{(6, 84)} = 2.00$, $p = .075$, $R^2 = .13$, though rotation time showed a negative effect ($\beta = -0.03$, $p = .024$). The Week 2 model was significant, $F_{(6, 86)} = 3.00$, $p = .010$, $R^2 = .17$, with rotation time emerging as a significant negative predictor of MoCA scores ($\beta = -0.06$, $p = .030$). The Week 3 model was also significant, $F_{(6, 86)} = 4.37$, $p < .001$, $R^2 = .23$. Here, being female predicted higher MoCA scores ($\beta = -0.92$, $p = .045$), and both longer rotation time ($\beta = -0.02$, $p = .031$) and higher Perspective Taking error were associated with lower MoCA scores.

Table 3. Linear regression models predicting MoCA scores. Panel A includes SPACE error indices by week. Panel B includes task-specific predictors (Movement time, Rotation time, Perspective Taking error).

### A: Models with SPACE error indices

|  | Baseline | Week 1 | Week 2 | Week 3 |
|---|---|---|---|---|
| (Intercept) | 27.38*** (0.88) | 26.56*** (1.00) | 26.20*** (0.93) | 26.48*** (0.86) |
| Age | −0.02 (0.01) | −0.00 (0.02) | 0.01 (0.02) | 0.01 (0.02) |
| Gender [Male − Female] | −0.64 (0.48) | −0.65 (0.48) | −0.78 (0.46) | −0.99* (0.46) |
| Education [University − High School] | 0.66 (0.52) | 0.62 (0.52) | 0.24 (0.52) | 0.15 (0.50) |
| SPACE error Week 1 | — | −0.51 (0.38) | — | — |
| SPACE error Week 2 | — | — | −1.07** (0.36) | — |
| SPACE error Week 3 | — | — | — | −1.24*** (0.33) |

| **Model comparisons (A)** | | | | |
|---|---|---|---|---|
| Comparison | $\Delta R^2$ | $F$ | $(df_1, df_2)$ | $p$ |
| Baseline vs Model 1 | 0.02 | 1.85 | (1, 87) | .177 |
| Baseline vs Model 2 | 0.08 | 8.79 | (1, 88) | .004 |
| Baseline vs Model 3 | 0.13 | 14.01 | (1, 88) | <.001 |

### B: Models with task-specific predictors

|  | Baseline | Week 1 | Week 2 | Week 3 |
|---|---|---|---|---|
| (Intercept) | 27.38*** (0.88) | 27.73*** (1.16) | 29.53*** (1.26) | 29.12*** (1.16) |
| Age | −0.02 (0.01) | −0.01 (0.02) | 0.01 (0.02) | −0.00 (0.02) |
| Gender [Male − Female] | −0.64 (0.48) | −0.69 (0.48) | −0.75 (0.46) | −0.92* (0.45) |
| Education [University − High School] | 0.66 (0.52) | 0.49 (0.52) | 0.33 (0.53) | 0.12 (0.50) |
| Movement Time | — | 0.01 (0.02) | −0.03 (0.03) | 0.01 (0.01) |
| Rotation Time | — | −0.03* (0.01) | −0.06* (0.03) | −0.02* (0.01) |
| Perspective Taking Error | — | 0.00 (0.01) | −0.01 (0.01) | −0.03*** (0.01) |

| **Model comparisons (B)** | | | | |
|---|---|---|---|---|
| Comparison | $\Delta R^2$ | $F$ | $(df_1, df_2)$ | $p$ |
| Baseline vs Model 1 | 0.06 | 1.95 | (3, 84) | .128 |
| Baseline vs Model 2 | 0.10 | 3.55 | (3, 86) | .018 |
| Baseline vs Model 3 | 0.16 | 6.09 | (3, 86) | <.001 |

**Notes.** Entries are coefficients with standard errors in parentheses. Stars denote significance: * $p < .05$, ** $p < .01$, *** $p < .001$. Baseline includes age, gender, education only. Panel A: SPACE error indices added by week. Panel B: task-specific predictors (movement, rotation, Perspective Taking) added by week.

### 4.3　Q3: Usability

To evaluate whether usability scores met established benchmarks, we conducted one-sample Wilcoxon signed-rank tests against the respective reference values. For the SUS (benchmark = 68), Week 1 scores did not differ significantly from the benchmark ($Z = -1.33$, $p = .18$, $r = -0.14$). In contrast, Weeks 2 and 3 were significantly higher, Week 2 ($Z = -3.59$, $p < .001$, $r = -0.37$); Week 3 ($Z = -3.70$, $p < .001$, $r = -0.39$). For the NASA-TLX (benchmark = 50), scores were significantly lower than the midpoint for Week 1 ($Z = -7.93$, $p < .001$, $r = -0.83$), Week 2 ($Z = -8.18$, $p < .001$, $r = -0.85$), and Week 3 ($Z = -8.09$, $p < .001$, $r = -0.84$). For the UEQ (benchmark = 0), attractiveness scores did not differ significantly from neutrality across all weeks (all $p > .05$). In contrast, Pragmatic quality was significantly above zero for Week 1 ($Z = 2.82$, $p < .001$, $r = 0.29$), Week 2 ($Z = 4.00$, $p < .001$, $r = 0.41$), and Week 3 ($Z = 2.69$, $p < .001$, $r = 0.28$). Hedonic quality was significantly below zero for Week 1 ($Z = -6.17$, $p < .001$, $r = 0.64$), Week 2 ($Z = -7.05$, $p < .001$, $r = 0.73$), and Week 3 ($Z = -7.55$, $p < .001$, $r = 0.78$).

*4.3.1　Q3.1: Usability changes over Week.* Results of the ART ANOVA with the SUS scores revealed a significant main effect of Week, $F(2, 194) = 9.62$, $p < .001$, $W = 0.10$. Post hoc comparisons indicated that scores in Week 2 ($p = .001$, $r_{rb} = 0.38$) and Week 3 ($p < .001$, $r_{rb} = 0.40$) were significantly higher than Week 1, while Weeks 2 and 3 did not differ ($p = .752$). For the UEQ subscales, Attractiveness and Pragmatic quality showed no significant week effects (all $p > .59$). In contrast, Hedonic quality yielded a significant week effect, $F(2, 194) = 8.68$, $p < .001$, $W = 0.04$ with higher ratings in Week 2 ($p = .002$, $r_{rb} = -0.22$) and Week 3 ($p < .001$, $r_{rb} = -0.32$) compared to Week 1, while Weeks 2 and 3 did not differ. Finally, NASA-TLX scores also showed a significant week effect, $F(2, 182) = 6.39$, $p = .002$. Post hoc comparisons revealed that the cognitive workload was lower in Week 2 than in Week 1 ($p = .001$, $r_{rb} = -0.39$), while Week 3 did not significantly differ from either Week 1 ($p = .086$) or Week 2 ($p = .131$).

### 4.4　Q4: Effects of Gender, Age Group, and Week on SPACE Performance

Results of a 2 (Gender) × 3 (Age Group) × 3 (Week) ART ANOVA with SPACE error as the dependent variable revealed significant main effects of Age Group, $F(2, 87) = 22.43$, $p < .001$, $\epsilon^2 = 0.26$, and Week, $F(2, 174) = 131.28$, $p < .001$, $W = 0.61$, but no significant main effect of Gender, $F(1, 87) = 2.35$, $p = .129$. For age group, post hoc tests revealed that younger adults ($M = 139$) performed significantly better (lower error) than middle-aged adults, $t(87) = -3.13$, $p = .007$, $\delta = -0.36$, and older adults, $t(87) = -6.69$, $p < .001$, $\delta = 0.80$, while middle-aged adults also outperformed older adults, $t(87) = -3.57$, $p = .002$, $\delta = -0.54$. For the main effect of week, post hoc tests revealed that performance improved from Week 1 ($M = 125.6$) to Week 2 ($M = 91.3$), $t(174) = 4.86$, $p < .001$, $\delta = 0.28$, but errors increased sharply in Week 3 ($M = 203.0$), which was significantly higher than both Week 1, $t(174) = -10.95$, $p < .001$, $\delta = 0.59$, and Week 2, $t(174) = -15.82$, $p < .001$, $\delta = 0.75$. The analysis revealed no significant two- or three-way interactions (all $p > .054$).

### 4.5　Q5: Effects of Gender, Supervision, and Age Group on SPACE Performance

Results of a 2 (Gender) × 2 (Supervision) × 3 (Age Group) ART ANOVA with SPACE error as the dependent variable revealed significant main effects of Gender, $F(1, 1070) = 7.91$, $p = .005$, $\delta = 0.14$, Age Group, $F(2, 1070) = 20.18$, $p < .001$, $\epsilon^2 = 0.06$, and Supervision, $F(1, 1070) = 11.59$, $p < .001$, $\delta = 0.24$. The Age Group × Supervision interaction was also significant, $F(2, 1070) = 3.37$, $p = .035$. No other two-way or three-way interactions were significant (all $p > .10$). Post hoc analyses revealed that females ($M = 589$) had significantly higher SPACE error scores than males ($M = 482$), $t(1070) = 2.81$, $p = .005$, $\delta = 0.14$. Younger adults ($M = 287$) performed significantly better than both middle-aged ($M = 459$), $t(1070) = -3.53$, $p = .001$, $\delta = -0.35$, and older adults ($M = 571$), $t(1070) = -6.35$, $p < .001$, $\delta = -0.51$, and middle-aged participants performed better than older adults, $t(1070) = -2.59$, $p = .026$, $\delta = -0.21$. Participants in the supervised condition ($M = 551$) also showed higher error than those in the unsupervised condition ($M = 421$), $t(1070) = 3.41$, $p = .001$, $\delta = 0.24$. Regarding the Age Group × Supervision interaction, follow-up contrasts revealed that the overall supervision effect was driven by older adults. Specifically, older adults in the unsupervised condition ($M = 746$) had significantly higher SPACE error scores than those in the supervised condition ($M = 529$), $t(1070) = -3.82$, $p = .002$, $\delta = 0.21$. No supervision effects were observed for younger (20-40) or middle-aged (41-60) adults (all $p > .60$). Figure 6 illustrates these results.

## 4.6 Q6: Effects of Gender, Supervision, and Age Group on Usability and Cognitive Workload

Table 4. Demographic characteristics for the previous SPACE studies conducted under supervision.

| Study type | N | Female | Male | Age range | Medium age | Mean age | SD |
|---|---|---|---|---|---|---|---|
| uSPACE(Simplified) | 122 | 65 | 57 | 21–76 | 43 | 45.84 | 15.76 |
| bioSPACE | 298 | 174 | 124 | 53–89 | 74 | 74.23 | 6.46 |
| targetSPACE | 566 | 308 | 258 | 60–92 | 71 | 71.49 | 7.12 |
| trainSPACE | 40 | 0 | 40 | 55–79 | 67 | 66.65 | 5.98 |
| Overall | 1026 | 547 | 479 | 21–92 | 68 | 69.05 | 12.09 |

- **uSPACE:** A study by Colombo et al. that evaluated the usability of the full SPACE game across three experiments, examining different control interfaces, the cognitive load imposed by a UI element, and the overall complexity of different trial and landmark configurations. The combined simplified design, which incorporated semi-automated control, a rotation widget, and reduced task complexity, led to modest but consistent usability improvements across all age groups.[20] Another study by Colombo et al. focused on investigating the relationship between full SPACE and the MoCA in a cohort of 348 participants, showing that SPACE tasks, particularly perspective taking, significantly predicted MoCA scores beyond demographic risk factors.[22] The 122 participants in this table belong to the Simplified condition in the papers by Colombo et al. [20, 22]. This condition was chosen since it had the highest usability and and was reported as the standard in all later SPACE studies below.
- **bioSPACE:** This study validated full SPACE in a memory clinic and community cohort (n = 298), showing that it reliably discriminated across Clinical Dementia Rating stages with higher accuracy than models including demographic variables ([21].
- **targetSPACE:** The targetSPACE cohort consisted of 566 participants and assessed the association between the full SPACE battery and the MoCA in individuals aged 60 and above.
- **trainSPACE:** This study investigated whether performance in SPACE could predict hippocampal volume in a sample of 40 healthy older adults. Results revealed that SPACE performance, particularly path integration and mapping tasks, significantly predicted hippocampal volume beyond age, education, and standard neuropsychological tests[69].

Results of three 2 (Gender) × 2 (Supervision) × 3 (Age Group) ART ANOVAs with SUS revealed significant main effects of Gender, $F(1, 202) = 16.42$, $p < .001$, $\delta = -0.31$, Supervision, $F(1, 202) = 35.26$, $p < .001$, $\delta = -0.45$, and Age Group, $F(2, 202) = 13.43$, $p < .001$, $\epsilon^2 = -0.10$ , but no significant interactions (all $p \geq .275$). Participants in the supervised condition ($M = 86.2$) had significantly lower scores than those in the unsupervised condition ($M = 135.2$), $t(202) = -5.94$, $p < .001$, $\delta = -0.45$ . Younger adults ($M = 133$) reported significantly higher scores than both middle-aged ($M = 103$), $t(202) = 3.05$, $p = .007$ , $\delta = 0.29$, and older adults ($M = 79$), $t(202) = 5.13$, $p < .001$, $\delta = 0.46$, while middle-aged adults reported higher scores than older adults, $t(202) = 2.32$, $p = .056$, $\delta = 0.23$.

For UEQ Attractiveness, no main effects or interactions were detected (all $p \geq .120$). For UEQ Pragmatic, there was a significant main effect of Supervision, $F(1, 202) = 10.50$, $p = .001$, $\delta = -0.26$. Participants in the unsupervised condition ($M = 123.7$) reported significantly higher ratings than those in the supervised condition ($M = 95.4$), $t(202) = -3.24$, $p = .001$, $\delta = -0.26$. For UEQ Hedonic, there was a significant main effect of Age Group, $F(2, 202) = 9.04$, $p < .001$, $\epsilon^2 = 0.08$, with older adults ($M = 128.8$) reporting significantly higher scores than younger ($M = 85.2$), $t(202) = 115.1$, $p < .001$, $\delta = 0.39$, and middle-aged adults ($M = 115.1$), $t(202) = -3.01$, $p = .008$, $\delta = 0.15$. The difference between younger and middle-aged adults was not significant, $t(202) = -1.28$, $p = .407$, $\delta = -0.30$. Gender and Supervision were not significant (both $p > .06$).

For NASA-TLX, there were significant main effects of Gender, $F(1, 202) = 7.80$, $p = .006$, $\delta = 0.22$, Supervision, $F(1, 202) = 114.14$, $p < .001$, $\delta = 0.70$, and Age Group, $F(2, 202) = 7.98$, $p < .001$, $\epsilon^2 = 0.26$, with no significant interactions (all $p \geq .477$). Post hoc analyses revealed that females ($M = 119.2$) reported significantly higher cognitive workload than males ($M = 94.6$), $t(202) = 2.79$, $p = .006$, $\delta = 0.22$. Participants in the supervised condition ($M = 140.2$) had significantly higher cognitive workload than those in the unsupervised condition ($M = 64.2$), $t(202) = 10.68$, $p < .001$, $\delta = 0.70$. Younger adults ($M = 89.7$) reported significantly lower cognitive workload than middle-aged ($M = 106.4$), $t(202) = -1.67$, $p = .040$, $\delta = -0.17$, and older adults ($M = 132.9$), $t(202) = -3.99$, $p < .001$, $\delta = -0.38$, while middle-aged adults also reported lower cognitive workload than older adults, $t(202) = -2.45$, $p = .040$, $\delta = -0.26$. Besides the statistics, we additionally visualised the group means and standard errors using a line plot (Fig. 7).

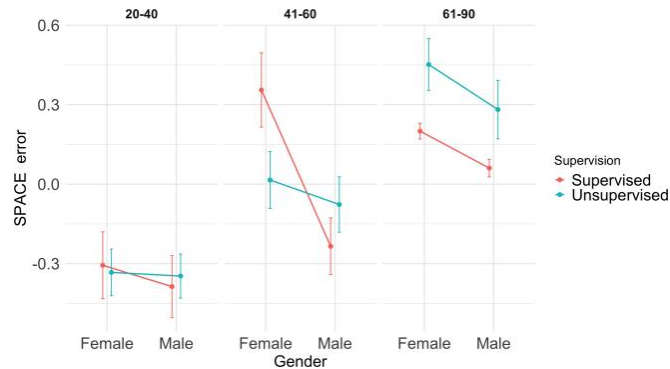

Fig. 6. Comparison of SPACE error across supervised and unsupervised conditions, plotted by gender and age group. Points show mean performance, and error bars represent standard errors.

## 4.7 Qualitative Analysis

To complement the quantitative usability metrics, we conducted a qualitative content analysis following the literature [58] of participants' open-ended responses collected after gameplay. Responses were first screened for relevance and then inductively coded by theme, with recurrent categories grouped into higher-order themes such as instruction clarity, usability issues, and engagement. Figure 8 presents a Sankey diagram visualising the flow frequency between the different age groups and gender across the identified themes.

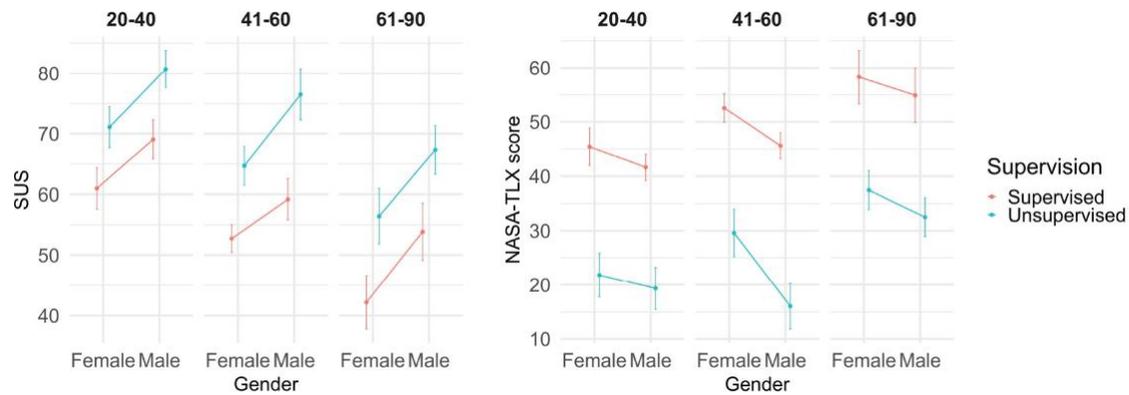

Fig. 7. SUS scores (left) and NASA-TLX (right) by supervision condition (supervised vs. unsupervised), gender, and age group. Points represent mean values, and error bars indicate standard errors.

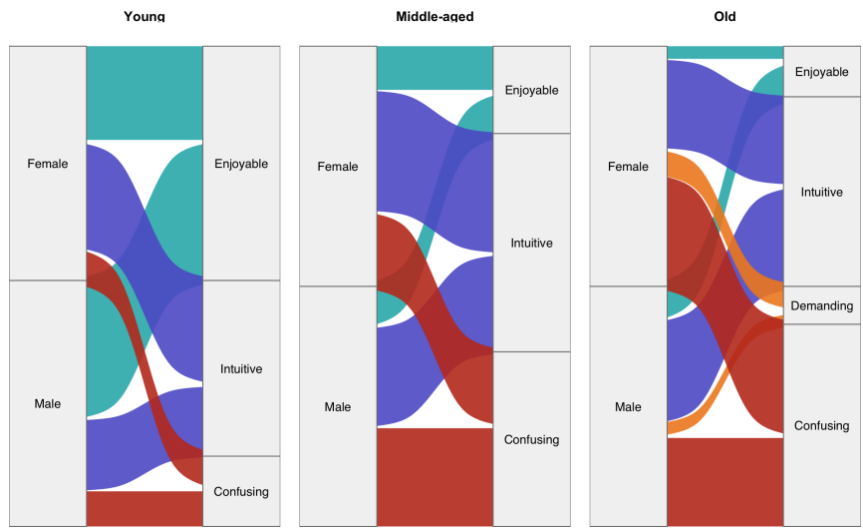

Fig. 8. Sankey diagrams showing participants' qualitative feedback themes across Age Group (Young, Middle-aged, Old) and Gender. Four themes emerged from open-ended responses: Enjoyable (blue), Intuitive (purple), Confusing (red), and Demanding (orange). The width of the flows represents the number of responses in each category, illustrating how perceptions of the game tasks varied by Age Group and Gender.

*4.7.1 Instruction clarity and usability.* In qualitative feedback, approximately 29.2% of the participants reported that the game instructions for both tasks were clear and the controls were easy to use, with representative comments such as *"I found the instructions and illustrations quite clear and easy to follow.".* Among those who perceived the instructions and controls as clear, 41% were female. As expected, younger (50%) and middle-aged participants (33%) found the instructions easier to use compared to older participants (17%). Critically, 63.4% of participants indicated that while the instructions for the Visuospatial Training task were clear, the objectives and instructions for the Perspective Taking task were less understandable. Here, representative comments included *"As for the [Perspective Taking task], it was not very clear on the objectives and instructions"* and *"Instructions are wordy and small in print."* Finally, a handful of participants noted other usability issues, such as unnecessarily long pauses between animated instructional segments, minor lags during interactions, and difficulties with icon visibility.

*4.7.2 Engagement.* Approximately 45% of participants expressed that the game could be made more engaging and enjoyable, with 44% of these respondents being older adults above 60 years of age. Several participants noted that the game contained repetitive trials and risked becoming monotonous if task difficulty did not increase over time. To address this, participants suggested gradually increasing the difficulty level, introducing additional obstacles or multiple landmarks, and incorporating more interactive features to maintain interest over repeated sessions. As one participant remarked, *"[m]aybe introduce levels or challenges that get progressively harder, or even add a leaderboard if it's competitive."* Others emphasised the value of richer visual and auditory stimulation, noting that mini-SPACE should include *"[...] more interactive games, with more audio stimulation to keep the elderly engaged".* A few participants also suggested cosmetic changes to improve immersion, such as *"mak[ing] the rocket icon really look like a rocket. Other than that, this is great!"*

## 5 Discussion

In this study, we evaluated the reliability, validity, and usability of mini-SPACE, a serious game designed to detect CI, when administered without supervision across a three-week period. First, we examined the test-retest reliability of mini-SPACE by tracking performance across three consecutive weeks (Q1). Second, we assessed validity relative to an established benchmark by relating performance on mini-SPACE to MoCA scores (Q2). Third, we examined usability under unsupervised conditions by comparing mini-SPACE scores in the SUS, UEQ, and NASA-TLX against established benchmarks (Q3). Here, we also assessed whether performance and usability ratings changed over time, testing for learning effects in task performance or shifts in subjective usability and cognitive workload ratings (Q3.1). Fourth, we examined whether performance varied across age, gender, and repeated testing sessions (Q4). Fifth, we compared performance in unsupervised mini-SPACE to previous supervised versions of SPACE to determine whether performance was consistent across testing contexts (Q5). Sixth, we compared usability ratings between supervised and unsupervised conditions to evaluate whether the unsupervised format maintained measurement robustness while increasing usability and reducing cognitive workload (Q6).

Spearman rank-order correlations consistently revealed strong positive associations ($\rho$ = .60 - .75) across weeks, indicating that performance rankings were preserved over time. According to established benchmarks, these correlations fall within the moderate-to-good reliability range. Notably, performance stability improved after the first week, suggesting that once participants became familiar with the game mechanics, their task execution became more consistent. The slightly weaker reliability between Week 1 and Week 3 likely reflects the combined influence of lower familiarity at the outset and the increased task demands introduced in Week 3. Given that the prodromal phase of dementia can span decades, reliable home-based unsupervised longitudinal assessments are essential, making test-retest reliability a critical requirement across target populations [1]. Indeed, a recent review of digital cognitive tools for detecting to detect CI highlighted that while most studies focus on the feasibility and cross-sectional validity of emerging tools, reliability is often under-reported [91] and varies widely between studies (ICCs -0.06 to 0.97) [79]. By demonstrating strong test-retest reliability for mini-SPACE in a fully unsupervised, home-based context, our findings directly address this gap and position it as a tool for the longitudinal monitoring of cognitive health.

Regarding changes in overall performance across the three weeks, we found that errors decreased from the first to the second week, likely reflecting a short-term learning effect as participants familiarised themselves with the task. As expected, error rates rose when task complexity and trial number were increased, suggesting that mini-SPACE is sensitive to capturing higher cognitive demands. Prior work has linked diminished practice effects to amyloid burden and heightened risk of subsequent CI [101]. By structuring Weeks 1 and 2 with an identical configuration and then increasing the difficulty in Week 3, mini-SPACE simultaneously allows for practice effects (or the lack of) to be observed as a potential marker [30, 35; but see 33] while reducing the risk that performance improvements are merely a reflection of task familiarity, making the assessment sensitive to subtle impairment that may emerge under more demanding conditions.

To evaluate the validity of mini-SPACE against traditional cognitive screening assessments, we related SPACE outcomes for each week to MoCA scores (Table 3). The MoCA is one of the most widely used instruments for detecting CI [72] and is often preferred to other paper-and-pencil assessments, such as MMSE [36], due to its superior sensitivity [54, 78, 92]. The MoCA also provides a cheaper alternative to full neuropsychological batteries and is less invasive than fluid and imaging biomarker assessments. Indeed, the MoCA has been frequently used in digital health research as a reliable, but not diagnostic, benchmark to validate new tools aimed at distinguishing between healthy ageing, CI and dementia [63, 75, 88].

Our findings revealed that performance in mini-SPACE significantly predicted MoCA scores beyond age, gender, and education. Importantly, as detailed in the supplemental material, SPACE performance also remained a significant predictor after controlling for a range of modifiable risk factors linked to dementia. These results are consistent with our earlier work [22] using the full and supervised version of SPACE, in which we also found that spatial and navigation performance, particularly the Perspective Taking task, explained variance in MoCA beyond demographic and risk factors. Similar to Colombo et al. [22], we believe that the reason a notable proportion of the variance in MoCA scores remains unexplained is that the spatial and navigation tasks in SPACE and mini-SPACE provide complementary information to traditional cognitive assessments by targeting medial temporal lobe dysfunction. Indeed, our most recent work [21] has shown that both the full battery of SPACE, as well as the shortened version consisting only of the Visuospatial Training and Perspective Taking tasks, predict with high sensitivity and specificity clinically diagnosed adjacent stages of CI as assessed by the Clinical Dementia Rating Scale (CDR) [70].

Looking more closely into the week-by-week patterns, we found that our models significantly predicted MoCA starting in Week 2 and showed stronger effects by Week 3. These results are further supported by the correlation analysis in Figure 5, showing that the unsupervised condition — particularly Week 3 — approached the strength of the MoCA associations observed in the supervised setting from previous studies [20, 22]. When considering which mini-SPACE task best predicted MoCA scores, we found that longer rotation times in Week 2 and Week 3 and higher Perspective Taking errors in Week 3 independently contributed to lower MoCA scores. This progression suggests that while rotation/object tracking ability is an early marker of CI, Perspective Taking becomes increasingly relevant and surpasses rotation once participants are familiar with the tasks. Given that each task captures complementary aspects of spatial cognition at different stages of engagement, we believe that combining them into a composite SPACE error score provides a more stable and sensitive index of CI than relying on either measure alone [91].

We evaluated the usability of SPACE by comparing usability scores against established benchmarks for the SUS, UEQ and NASA-TLX. We found that mini-SPACE is rated positively under unsupervised conditions, with SUS scores significantly exceeding the commonly accepted benchmark of 68 already in Week 1 [6][50]. Critically, SUS scores improved significantly from Week 1 to Week 2 and then remained stable. This pattern suggests that although usability was already satisfactory after the first exposure, participants benefited from additional familiarisation. Similarly, UEQ ratings of Attractiveness, Pragmatic quality, and Hedonic quality were all significantly above the neutral benchmark of zero [59], reflecting a favourable user experience. Here, UEQ Attractiveness scores did not differ significantly from neutrality across weeks, indicating that participants' overall impression of mini-SPACE's appeal remained stable and did not deteriorate with repeated use. Pragmatic quality was rated positively across all weeks, highlighting that participants consistently found the tool to be useful. In contrast, Hedonic quality was rated significantly below neutrality across all weeks, suggesting that participants did not experience the game to be especially stimulating. Taken together, these findings show that SPACE was perceived as usable in unsupervised settings, with stable impressions of its practical utility, but that improvements can be made regarding its capacity to stimulate.

Cognitive workload was generally very low, with NASA-TLX scores consistently falling below the 50 midpoint [40] and decreasing from Week 1 to Week 2, with Week 3 not differing significantly from either Week 1 or Week 2. Previous work has shown that SUS ratings tend to remain stable across repeated administrations over weeks to months [12]. In contrast, greater experience with a product has been found to be associated with higher SUS scores, suggesting that familiarisation can potentially raise perceived usability independent of design changes [68]. The fact that neither SUS nor cognitive workload decreased in Week 3, despite the increase in the number of trials and complexity of the tasks, suggests that once participants invested an initial effort and learned to interact with mini-SPACE, the added cognitive demands did not undermine the ease of use. This pattern mirrors our performance results relative to the MoCA, in which performance in mini-SPACE was more predictive of MoCA scores in Week 2 compared to Week 1, culminating in Week 3. Together, these results highlight an important usability lesson and suggest that familiarisation in the early stages is essential for providing a stable user experience and capturing optimal predictive validity.

The analysis of gender and age effects on SPACE performance revealed a clear pattern. Gender did not significantly influence error rates, but age emerged as a strong determinant of performance. This observed age effect is consistent with previous findings from both general cognitive [57, 93] and navigation-specific assessments [18, 61, 86, 97]. Across the three weeks, older adults showed higher SPACE error compared to younger and middle-aged participants, in line with the well-established effect of ageing on spatial navigation abilities. Performance also significantly varied across weeks. Here, errors decreased sharply after the first session, indicating rapid learning and task familiarisation with a similar configuration and trials between Weeks 1 and 2, but rose slightly in Week 3 when additional trials and task demands were introduced. This trajectory parallels the reliability patterns reported above and suggests that mini-SPACE is sensitive to both learning effects and heightened cognitive load.

Our analysis of the effects of supervision on SPACE performance revealed three key findings relative to age and gender (Figure 6). First, consistent with the age-related patterns reported above, younger adults outperform both middle-aged and older adults, while middle-aged participants performed better than older adults. Second, females exhibited higher error rates than males. This finding is consistent with a large body of literature on spatial cognition and navigation showing a general male advantage in navigation tasks [10, 71, 73]. Interestingly, this gender effect was not evident in the repeated measures analysis across weeks, possibly because of the smaller sample when not accounting for the participants in the supervised condition. Most importantly, we found a significant interaction between age and supervision. While younger and middle-aged adults performed similarly across conditions, older adults showed a marked disadvantage in the unsupervised condition. Taken together, these findings suggest that while unsupervised testing is feasible and scalable, its implementation in older populations requires careful consideration to minimise error inflation arising from instruction-related misunderstandings rather than true cognitive deficits.

Finally, our analysis of the effect of supervision on usability and cognitive workload revealed that usability was consistently higher and cognitive load lower when SPACE was administered in unsupervised settings (Figure 7). Specifically, SUS and UEQ Pragmatic ratings indicated that participants found the interaction more efficient and goal-oriented and had an overall easier time interacting with mini-SPACE independently, while NASA-TLX scores revealed marked reduced workload in the absence of supervision. However, these results should be interpreted with some caution since the supervised studies employed the full battery of SPACE tasks, which requires more time to complete and is inherently more complex. Indeed, the previous work [20] has shown that mean SUS scores for the full SPACE battery fall below benchmark levels for older adults. Nevertheless, the fact that unsupervised mini-SPACE demonstrated strong usability while maintaining similar capacity to detect CI relative to the MoCA, indicates that a shorter, more targeted assessment may be particularly well suited for large-scale, home-based deployment.

Gender and age also emerged as significant factors influencing usability. With respect to gender, females reported lower SUS and higher cognitive workload scores than males. These results align with prior research on gender-related acceptance of technology [39]. Work by Sobieraj and Krämer [84] found that women tend to report lower technology self-efficacy and greater computer anxiety, especially when faced with complex systems. Critically, a recent meta-analysis [64] found that women tend to hold less positive attitudes towards digital games overall. Notably, the magnitude of the gender attitude disparity was higher for entertainment games than for serious games. The latter finding is encouraging because it suggests that both men and women see value in serious games, and future work should aim to further strengthen attitudes and co-develop tools with women to better meet their needs.

With respect to age, younger adults consistently reported the highest usability and lowest cognitive workload, followed by the middle-aged group and the older adults. These results are consistent with usability research showing that older adults perceive digital tools as less usable and more cognitively demanding [13, 29, 66, 85]. Interestingly, Hedonic ratings increased with age, with older participants reporting higher enjoyment despite higher error rates and cognitive workload. This is in line with our previous work in SPACE [20] and suggests that the tasks may have felt relatively simple and repetitive for younger participants. In contrast, the same level of difficulty may have been perceived as more novel and stimulating for older adults, many of whom have less gaming experience.

The qualitative analysis (Figure 8) further complements these results. Across all age groups, participants frequently described the game as enjoyable and intuitive, consistent with the generally high usability scores observed. Negative themes of confusion were more frequently reported by middle-aged and older adults, while comments on the demanding nature of the tasks came predominantly from older participants, particularly older females. In contrast, younger adults' feedback was dominated by positive themes. This pattern contrasts with findings from a previous study with the full SPACE version [20], where negative themes were more frequently reported across all three age groups. Altogether, these results indicate that mini-SPACE reduced some of the barriers for younger and middle-aged users, but that additional interface support and adaptive tutorials are needed for older adults.

## 6 Limitations

Although mini-SPACE was generally well received, some older adults reported relatively lower usability and higher cognitive workload, particularly during the first week. These remarks highlight that the initial familiarisation phase with novel technology can pose challenges for this age group which often lacks gaming experience [42]. To address this, future versions of mini-SPACE could integrate enhanced onboarding strategies, such as animated tutorials with synchronised voice-over instructions, which have been shown to improve comprehension and engagement among older users [56]. In addition, while the final number of participants was sufficient to validate the mini-SPACE's usability and assess its ability to predict CI, it is not sufficiently large for establishing robust age norms vis-à-vis performance and usability. Future work will therefore focus on recruiting a larger and more diverse participant pool to build a comprehensive database. This database will then be used to generate personalised game reports that quantify risk and present individuals with their performance relative to their gender and age group. A further limitation is that this study included healthy participants from the general population who were screened with the MoCA, preventing us from evaluating the sensitivity and specificity directly against clinical benchmarks. Future work will extend mini-SPACE to clinically diagnosed groups and benchmark performance against scales such as the CDR.

## 7 Conclusion

This study evaluated the reliability, validity, and usability of mini-SPACE, a short iPad-based serious game designed for large-scale, home-based cognitive screening. Across three weeks of testing, mini-SPACE showed strong test-retest reliability, with performance stability improving after the first week as participants became familiar with the game mechanics. While test-retest reliability was slightly weaker between Weeks 1 and 3, the result likely reflects a combination of lower familiarity at the outset and the increased task demands introduced in Week 3. With regard to validity, mini-SPACE significantly predicted MoCA scores beyond demographic variables and health variables, supporting its value as a digital marker of cognitive status. Importantly, predictive strength increased over the three weeks as participants grew more familiar with the tasks and the tasks themselves increased in complexity. As expected, performance and usability varied by age and gender: older adults, and to a lesser extent, females, generally performed worse and reported lower usability than middle-aged and younger adults. While all participants completed mini-SPACE unsupervised, older adults still struggled more without guidance. This challenge cannot be taken for granted, and continuously addressing this technology gap will be essential to ensure that mini-SPACE provides an equitable and accurate assessment across various age groups. Overall, our results demonstrate that mini-SPACE provides a reliable, valid, and usable tool for unsupervised digital cognitive assessment.

## 8 Acknowledgments